\pdfoutput=1
\documentclass{sigchi}

\CopyrightYear{2020}
\setcopyright{acmlicensed}
\conferenceinfo{CHI'20,}{April  25--30, 2020, Honolulu, HI, USA}
\acmPrice{\$15.00}
\doi{10.1145/3313831.3376782}
\isbn{978-1-4503-6708-0/20/04}

\usepackage{balance}       
\usepackage{graphics}      
\usepackage[T1]{fontenc}   
\usepackage{txfonts}
\usepackage{mathptmx}
\usepackage{color}
\usepackage{booktabs}
\usepackage{textcomp}
\usepackage{cite}
\usepackage{fontawesome}
\usepackage{multirow}
\usepackage[table,xcdraw]{xcolor}
\usepackage{rotating}
\usepackage{wrapfig}
\usepackage{url}

\usepackage[caption=false]{subfig}
\usepackage{comment}
\usepackage{tabularx}
\usepackage[normalem]{ulem}
\usepackage{calc}
\usepackage{enumitem}

\newcommand{\edit}[1]{\textcolor{black}{#1}}
\newcommand{\remove}[1]{\textcolor{purple}{\sout{}}}

\definecolor{sParam}{rgb}{0.502, 0.796, 0.769}
\definecolor{sDirectI}{rgb}{.565,.792,.976}
\definecolor{sIndirectI}{rgb}{.94,.71,.73}
\definecolor{sKeywords}{rgb}{1,.627,0}

\newcommand{\egP}[1]{\textcolor{sParam}{#1}}
\newcommand{\egDI}[1]{\textcolor{sDirectI}{#1}}

\newcommand{\egK}[1]{\textcolor{sKeywords}{#1}}

\newcommand{\symbolDI}{\raisebox{-.1em}{\includegraphics[height=.9em]{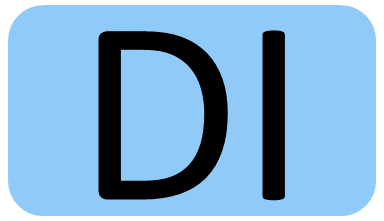}}}
\newcommand{\symbolP}{\raisebox{-.1em}{\includegraphics[height=.9em]{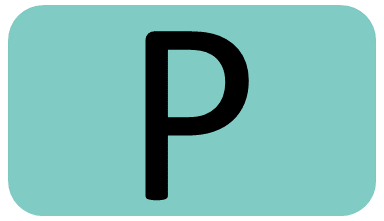}}}
\newcommand{\symbolK}{\raisebox{-.1em}{\includegraphics[height=.9em]{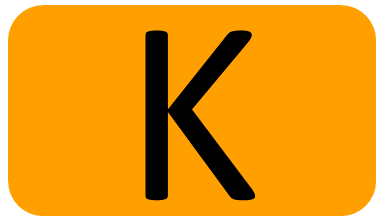}}}

\usepackage{microtype}        
\usepackage{ccicons}          

\usepackage{todonotes}

\def\plaintitle{InChorus: Designing Consistent Multimodal Interactions \\for Data Visualization on Tablet Devices}

\def\plainkeywords{Multimodal interaction; data visualization; tablet devices; pen; touch; speech.}

\makeatletter
\def\url@leostyle{%
  \@ifundefined{selectfont}{
    \def\UrlFont{\sf}
  }{
    \def\UrlFont{\small\bf\ttfamily}
  }}
\makeatother
\def\pprw{8.5in}
\def\pprh{11in}

\definecolor{linkColor}{RGB}{6,125,233}

\begin{document}

\title{\plaintitle}

\numberofauthors{1}
\author{
\begin{tabular}{c}
\hspace{-4mm}
Arjun Srinivasan$^{1, 2}$
\hspace{2mm}
Bongshin Lee$^{1}$
\hspace{2mm}
Nathalie Henry Riche$^{1}$
\hspace{2mm}
Steven M. Drucker$^{1}$
\hspace{2mm}
Ken Hinckley$^{1}$
\end{tabular}\\
\\
\begin{tabular}{cc}
\affaddr{$^{1}$Microsoft Research}
&
\affaddr{$^{2}$Georgia Institute of Technology}
\\
\affaddr{Redmond, WA}
&
\affaddr{Atlanta, GA}
\\
\affaddr{\normalsize{\{bongshin, nath, sdrucker, kenh\}@microsoft.com}}
& 
\affaddr{\normalsize{arjun010@gatech.edu}}
\\
\end{tabular}
}

\maketitle
\begin{abstract}
While tablet devices are a promising platform for data visualization, supporting consistent interactions across different types of visualizations on tablets remains an open challenge. In this paper, we present multimodal interactions that function consistently across different visualizations, supporting common operations during visual data analysis. By considering standard interface elements (e.g., axes, marks) and grounding our design in a set of core concepts including operations, parameters, targets, and instruments, we systematically develop interactions applicable to different visualization types. To exemplify how the proposed interactions collectively facilitate data exploration, we employ them in a tablet-based system, \textit{InChorus} that supports pen, touch, and speech input. Based on a study with 12 participants performing replication and fact-checking tasks with InChorus, we discuss how participants adapted to using multimodal input and highlight considerations for future multimodal visualization systems.
\end{abstract}




\section{Introduction}

Recent advancements in screen resolution and computing capabilities have made tablet devices a promising platform for data visualization.
With this potential in mind, numerous research projects have been investigating visualization tools on tablets (e.g.,~\cite{baur2012touchwave,drucker2013touchviz,sadana2014designing,rzeszotarski2014kinetica,jo2015wordleplus,sadana2016designing,sadana2016expanding,jo2017touchpivot}), facilitating interaction through touch and/or pen input.
Although these carefully designed systems highlight the potential of tablets along with key design considerations, two fundamental issues persist.

First, the majority of prior research about data visualization on tablets~\cite{baur2012touchwave,drucker2013touchviz,sadana2014designing,rzeszotarski2014kinetica,jo2015wordleplus} have focused on a single visualization type, optimizing interactions for that chart type.
This local optimization could result in a \textit{globally inconsistent interaction experience} when multiple types of visualizations are included as part of one system.
For example, prior systems have used the gesture of dragging a finger along the axis to sort a bar chart~\cite{drucker2013touchviz} and select points in a scatterplot~\cite{sadana2014designing}.
However, when both bar charts and scatterplots are supported by the same system, the gesture of dragging along an axis causes a conflict, resulting in inconsistent functionality across visualizations~\cite{sadana2016designing}.
Resolving such inconsistencies often requires system designers to introduce specialized gestures such as holding on the axis of a bar chart to enter a transient ``sort mode'' in which one can swipe to sort~\cite{sadana2016designing}.
Such subtle differences in gestures can be difficult to remember and may lead to errors while performing tasks, however.

Second, when depending only on pen and/or touch, systems face \textit{increased reliance on menus and widgets} as the number and complexity of operations grow.
For example, to filter, users have to select visual marks and tap a delete/keep-only button or adjust sliders and dropdown menus in control panels~\cite{sadana2014designing,sadana2016designing,jo2017touchpivot}.
Such indirect interactions with interface elements external to the objects of interest (e.g., marks) can divert the users' attention which may prove disruptive to their workflow~\cite{drucker2013touchviz}.
Additionally, given the space constraints of tablets, control panels can occlude the visualization and limit the screen space available for the visualization itself.

To address these issues, we propose using multimodal interactions where the directness and precision of pen and touch is complemented by the freedom of expression afforded by speech.
For example, 
we let users perform zoom \& pan through familiar pinch and drag touch gestures.
On the other hand, given its affordance for drawing free-form strokes, we offload actions like drawing a selection lasso to the pen.
Furthermore, considering its expressiveness, we use speech to support operations like filtering where one can filter selected points (e.g., ``\textit{Remove these}'' or ``\textit{Exclude others}'') or filter points satisfying specific criteria (e.g., ``\textit{Remove all movies except action, adventure, and comedy}'').

To design multimodal interactions that function consistently across different types of visualizations, we first surveyed \edit{18} visualization systems to identify common operations and interactions supported in current systems.
Then, by considering standard elements in visualization tools (e.g., axes, marks) and grounding our design in a set of core concepts including operations, parameters, targets, and instruments, we systematically develop interactions applicable to different visualization types.
To illustrate how the proposed interactions collectively facilitate visual data exploration, we employ them in a tablet-based system, \textit{InChorus}.
We also leverage InChorus to conduct a user study to assess the practical viability of the proposed interactions and observe if participants adapt to using multimodal input during common visual analysis tasks.
Based on a study with 12 participants, we found that participants successfully adapted to using the proposed interactions to complete a series of replication and fact-checking tasks, commenting favorably on the freedom of expression provided by multiple modalities.
Reflecting on our experience of designing interactions in InChorus and observations from the user study, we highlight promising research directions for future work on multimodal visualization systems.

In summary, we make the following contributions:
\vspace{-1mm}
\begin{itemize}[itemsep=.1\baselineskip]
    \item We present systematically designed multimodal interactions that function consistently across different types of visualizations, supporting core visual data analysis operations.
    \item Through a prototype system, InChorus, we exemplify how pen, touch, and speech-based multimodal interactions can collectively facilitate visual data exploration on tablets.
    \item We report findings from a user study, highlighting how multimodal input accommodates varying user interaction patterns and preferences during visual analysis.
\end{itemize}

\section{Related Work}

\subsection{Pen and Touch-based Visualization Systems}


A plethora of systems have investigated the use of touch and/or pen input for interacting with data visualization systems (e.g.,~\cite{lee2013sketchstory,lee2015sketchinsight,jo2015wordleplus,zgraggen2014panoramicdata,jo2017touchpivot,romat2019activeink,browne2011data,schmidt2010set,baur2012touchwave,drucker2013touchviz,rzeszotarski2014kinetica,sadana2014designing,thompson2018tangraphe}), examining different devices and form-factors including tablets (e.g.,~\cite{baur2012touchwave,drucker2013touchviz,sadana2014designing,jo2017touchpivot}), tabletops (e.g.,~\cite{isenberg2009collaborative,frisch2009investigating}), and large vertical displays (e.g.,~\cite{lee2015sketchinsight,zgraggen2014panoramicdata,lee2013sketchstory}), among others.

Although we consider several of these systems when designing multimodal interactions, most relevant to our work are prior systems that are designed for tablets~\cite{baur2012touchwave,drucker2013touchviz,sadana2014designing,sadana2016designing,jo2017touchpivot,rzeszotarski2014kinetica}.
For instance, with TouchWave~\cite{baur2012touchwave}, Baur et al.~presented a set of multi-touch gestures to interact with hierarchical stacked graphs on tablets, specifically noting that designing a consistent interaction set was one of the primary challenges they faced.
Drucker et al.~\cite{drucker2013touchviz} compared a gesture-based interface to a WIMP-based interface in the context of bar charts.
Their results showed that not only were people faster and more accurate with gestures, but also that people subjectively preferred direct, gestural interactions over interacting with controls in the WIMP interface.
Sadana and Stasko~\cite{sadana2014designing} presented multi-touch interactions for common operations including selections, zooming, and filtering, among others in the context of scatterplots.
Following up their work on scatterplots, Sadana and Stasko expanded their system to include other types of visualizations and support multiple coordinated views~\cite{sadana2016designing}.
However, upon including additional visualizations, they encountered challenges due to inconsistencies in interactions across visualizations, calling for future systems to leverage standard gestures and support consistent interactions and feedback~\cite{sadana2016designing}.
While these systems focus on touch-only input, with TouchPivot~\cite{jo2017touchpivot}, Jo et al.~illustrated how pen and touch input can complement WIMP-style interface elements to help novices conduct visual data exploration on tablets.

Our work is motivated by these examples from prior work and addresses a common challenge faced by these systems: inconsistency in interactions for different operations and visualization types.
Specifically, we systematically analyze the operations and interactions supported by these systems to identify our target operations and initial set of interactions.
Furthermore, while most current systems are optimized for a single type of visualization, to ensure consistency and general applicability of the developed interactions, we design and test our interactions in the context of five popular visualization types as part of the same system.

\subsection{Speech-based Multimodal Visualization Systems}


Recently, there has been an influx of natural language interfaces (NLIs) for data visualizations~\cite{cox2001multi,sun2010articulate,gao2015datatone,setlur2016eviza,hoque2018applying,aurisano2016articulate2,kassel2018valletto,srinivasan2018orko,yu2019flowsense}.
Although these systems focus on NL as their primary mode of interaction, they acknowledge the need for multimodal interaction to support limitations of NL such as ambiguity~\cite{gao2015datatone,setlur2016eviza}.
Furthermore, a majority of current NLIs explore the use of typed NL input in a desktop-setting.
However, typing is not an efficient input technique on interactive displays such as tablets where speech becomes a more natural form of input.
While the interpretation strategies may be comparable to typed NLIs, speech-based systems require different interface and interaction design considerations due to added complexity with potential speech-to-text recognition errors and the lack of assistive features such as auto-complete.

With this distinction between typed and spoken input in mind, two systems that are most related to our work are Orko~\cite{srinivasan2018orko} and Valetto~\cite{kassel2018valletto}.
Orko~\cite{srinivasan2018orko} supports multimodal touch- and speech-based interaction with node-link diagrams, facilitating common network visualization tasks including finding connections, computing paths, and attribute-based filtering of nodes, among others.
Valetto~\cite{kassel2018valletto} presents a conversational interface for people to query for visualizations using voice on tablets.
Once a visualization is created, Valetto allows performing a rotate gesture to flip X/Y axes and swipe to change the visualization type.
Although these systems support touch and speech input, they either focus on a specific visualization type~\cite{srinivasan2018orko} or conversational interaction~\cite{kassel2018valletto}, providing little insight into how the multimodal interactions were designed or how future systems can build upon the presented interactions.

In our work, we design pen, touch, and speech-based multimodal interactions for frequent visual analysis operations~\cite{heer2012interactive} enabling visual data exploration on tablets.
Furthermore, by leveraging common interface elements (e.g., axes, marks) and grounding our design in a set of core concepts applicable to most visualization systems, we also exemplify how future systems can build upon our work and systematically design and describe multimodal interactions with visualizations.
\section{Systematic Design of Multimodal Interactions for Data Visualization Tools}

\let\depth\relax

\begin{table*}[ht!]
    \centering
    \includegraphics[width=.95\linewidth]{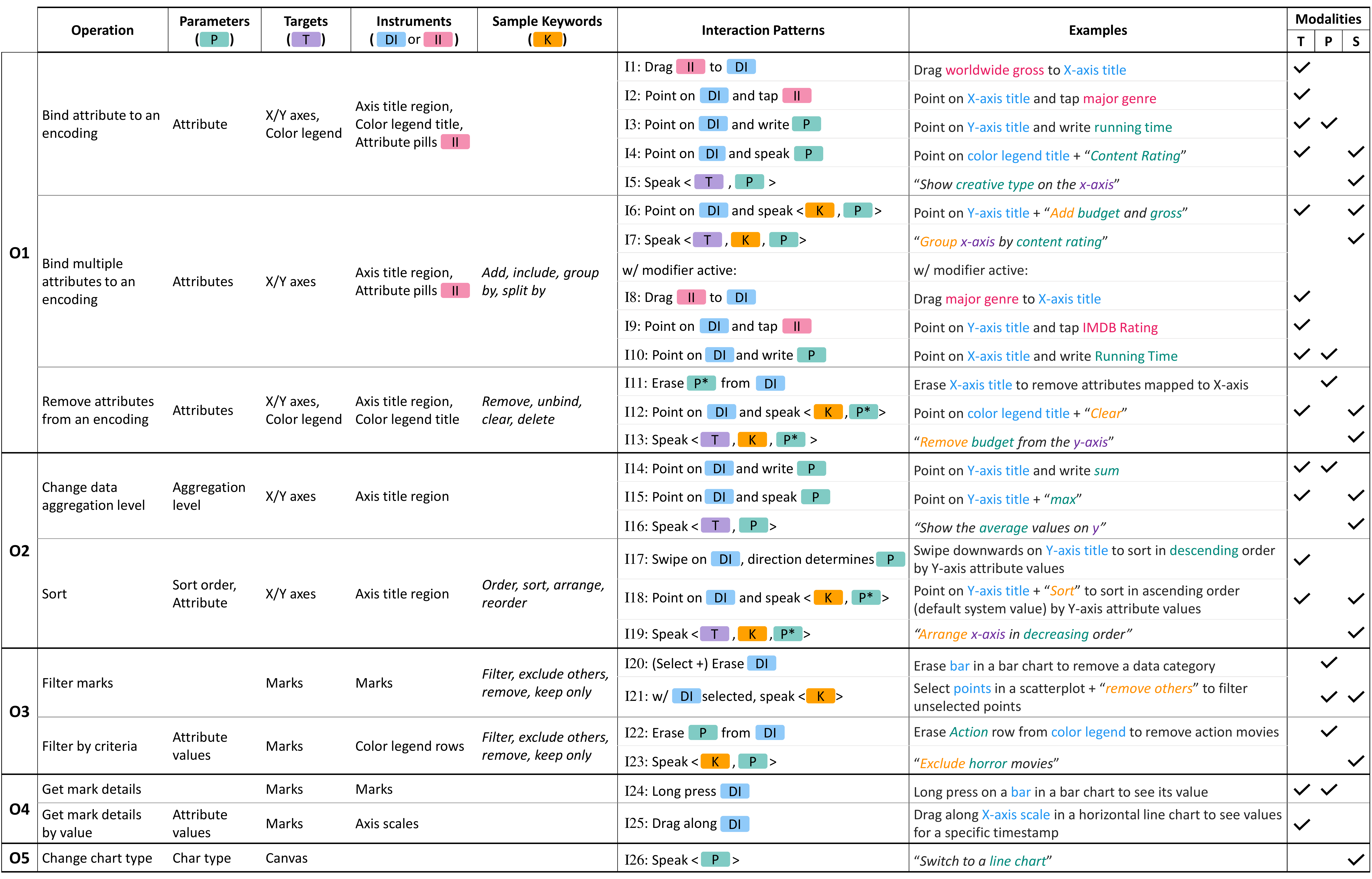}
    \caption{
    Proposed multimodal interactions for low-level operations during visual analysis.
    Operations categories are O1: Bind/unbind visual encodings, O2: Modify axes, O3: Filter, O4: Get details, and O5: Change chart type.
    Unless explicitly specified as an indirect instrument (II), all instruments are direct instruments (DI).
    An asterisk (*) indicates a parameter is optional.
    The rightmost column displays modalities (T: Touch, P: Pen, S: Speech) used in an interaction pattern.
    }
    \label{tbl:interactions}
\end{table*}

We surveyed \edit{18} visualization systems~\cite{subramonyam2018smartcues,walny2012understanding,kassel2018valletto,srinivasan2018orko,schmidt2010set,thompson2018tangraphe,jo2015wordleplus,jo2017touchpivot,zgraggen2014panoramicdata,rzeszotarski2014kinetica,sadana2016expanding,sadana2016designing,sadana2014designing,lee2015sketchinsight,browne2011data,baur2012touchwave,drucker2013touchviz,frisch2009investigating} to identify tasks and visualizations to consider as part of our design.
To explore a broader design space of multimodal interaction for data visualization, we started by considering pen, touch, and speech-based interactions with visualization tools regardless of the target device.
Specifically, our selection criteria were that systems (1) involved interactions using one or more of pen, touch, or speech input \edit{with a single device and user} and (2) focused on general visual data exploration and analysis, excluding systems that placed higher emphasis on externalizing users' thoughts (e.g.,~\cite{romat2019activeink,kim2019inking}) or authoring expressive visualizations (e.g.,~\cite{lee2013sketchstory,xia2018dataink,kim2019datatoon}).

Through the survey, we identified five core categories of operations (Table~\ref{tbl:interactions}) that were supported by most systems.
Furthermore, given their frequent occurrence in the surveyed systems and prevalence in common visualization tools, we decided to focus on histograms, bar charts (including grouped and stacked bar charts), line charts, scatterplots, and parallel coordinates plots as our initial visualization types.

\subsection{Conceptualizing Multimodal Interaction Design}
To design consistent interactions, we needed a standardized nomenclature to describe and compare alternative interactions.
Correspondingly, we reviewed the terminology and description of interactions in the surveyed visualization systems' papers.
However, since most current systems were optimized for specific visualizations (e.g.,~\cite{baur2012touchwave,drucker2013touchviz,thompson2018tangraphe,jo2015wordleplus,srinivasan2018orko}) or form-factors (e.g.,~\cite{lee2015sketchinsight,sadana2016designing,jo2017touchpivot,kassel2018valletto}),
there was no common language that let us consistently design and discuss possible interactions.
Thus, based on our survey and a review of prior work in the more general space of post-WIMP (e.g.,~\cite{van1997post,beaudouin2000instrumental}) and multimodal interfaces (e.g.,~\cite{cohen1997quickset-2,martin1998tycoon,oviatt2017handbook,srinivasan2019discovering}), we identified a set of core concepts that could help us (and future system designers) systematically design and reason about interactions in the context of multimodal visualization systems.

We propose thinking of interactions with multimodal visualization systems in terms of the following four concepts.
People interact with visualization systems through a set of one or more low-level \textbf{\textit{operations}} (e.g., binding an attribute to an encoding, sorting) to accomplish their high-level tasks (e.g., answering data-driven questions, creating specific visualizations).
These operations typically require \textbf{\textit{parameters}} (e.g., sorting order, attributes, encodings) and operate on one or more \textbf{\textit{targets}} (e.g., selected marks, axis, canvas).
Finally, operations are mediated through \textbf{\textit{instruments}} in the interface (e.g., marks, axes scales).
These instruments can be \textit{direct} (i.e., when the target itself mediates an operation) or \textit{indirect} (i.e., when an operation is performed on a target through a separate instrument).

With these \edit{general user interface} concepts in mind, we investigated possible interactions for the operations identified through the survey.
To ensure the resulting interactions were generalizable, we only considered basic elements (e.g., axes, marks, attribute pills) present in most visualization tools as our instruments.
We then examined both interactions demonstrated in previous systems (e.g., writing an aggregation function name to change the data aggregation level~\cite{jo2017touchpivot}, dragging along an axis to sort~\cite{drucker2013touchviz}) as well as novel interactions
\edit{that were potentially more fluid and consistent}
(e.g., pointing on an axis with a finger and speaking an attribute name to specify mappings, using the pen's eraser to filter).

\edit{Table~\ref{tbl:interactions} lists the ten low-level operations and the corresponding set of interactions (\textit{I1-I26}) derived after a series of iterations, along with examples and input modalities.}
We initially considered \textit{selection} and \textit{zoom/pan} as two additional core operations.
However, since these are low-level interface actions and sometimes a precursor to other operations (e.g., filtering a set of marks may require selecting them first), we decided not to include them as standalone operations.
In our interaction set, selections can be performed in four ways: 1) tapping with pen/finger directly on a mark, 2) tapping with pen/finger on a legend item to select marks by categories~\cite{heer2008generalized}, 3) dragging the pen on an axis scale to select marks based on data values, and 4) drawing a free-form lasso with the pen on the chart area.
Additionally, zoom and pan are supported through the standard two-finger pinch and single finger drag gestures on the chart area, respectively.

\subsection{Design Principles}
In this section, we describe five underlying principles we had when designing our multimodal interactions.
We compiled these principles based on the surveyed papers as well as design guidelines from prior work advocating for post-WIMP visualization interfaces~\cite{elmqvist2011fluid,lee2012beyond,roberts2014visualization}.

\subsubsection{\textbf{DP1.} Maintain interaction consistency across visualizations}

To support consistency, we prioritize globally functional interactions (i.e., ones that work across visualization types) over locally optimal interactions (i.e., ones that are specific to a type of visualization).
A pattern from Table~\ref{tbl:interactions} exemplifying this principle is \textit{I15: Drag along axis scales} to see mark details. 
Previous systems have inconsistently used this interaction to sort bars in a bar chart~\cite{drucker2013touchviz} and select points in a scatterplot~\cite{sadana2014designing}.
However, since these are both locally optimal interactions, we use dragging along an axis scale to display mark details which is a common operation across visualizations.


We note that some operations are specific to certain visualization types.
For example, sorting an axis is meaningful to bar charts and parallel coordinate plots but not to scatterplots and line charts.
Thus, \textit{I17:swiping on the axes} only works for the appropriate visualizations and has no effect in others.
We also reserve the swipe interaction for sorting and do not employ it for a different operation elsewhere.


\subsubsection{\textbf{DP2.} Minimize indirection in interactions}

Aligned with the guidelines for fluid interaction~\cite{elmqvist2011fluid}, we try to enable interactions with direct instruments (e.g., marks, axes), avoiding external controls and indirect instruments that are separated from the view.
For instance, to filter marks, people can use \textit{I20: erase marks} directly instead of adjusting external widgets like sliders or dropdown menus.
Or to see the details of a mark, one can use \textit{I24: long press on marks} instead of indirectly requesting for details through voice.

Another implication of this design principle is that if an operation is inherently indirect, we offload it to speech since it is also, by nature, indirect.
Examples of such indirect operations include changing the visualization type and filtering based on attributes that are not encoded in the current view (e.g., filtering points by \textit{imdb rating} in a scatterplot of \textit{production budget} by \textit{worldwide gross}).

\subsubsection{\textbf{DP3.} Leverage simple and familiar gestures}

Simple gestures that are familiar to users are easier to learn and subjectively preferred for interacting with pen and touch-based visualization systems~\cite{walny2012understanding,drucker2013touchviz,sadana2016designing,jo2017touchpivot}.
To maintain simplicity and promote familiarity, we avoid devising specialized gestures for individual operations.
In fact, as illustrated by the patterns in Table~\ref{tbl:interactions}, all our pen and touch interactions only involve common gestures including tap, point (hold), swipe, and drag.
Particularly for cases where one gesture could be mapped to multiple operations, instead of introducing an alternative gesture for one of the operations, we apply a division of labor tactic~\cite{hinckley2010pen+} and offload the interaction to a different modality.
For example, due to their ubiquity across devices and applications, we reserve touch-based pinch and drag gestures for zoom and pan, respectively.
However, dragging on the chart area is also an intuitive way to perform selection (e.g., by drawing lassos~\cite{sadana2014designing}), which is another important action during visual analysis~\cite{yi2007toward}.
To resolve this conflict, we leverage a second modality and allow people to draw selection lassos by dragging on the chart area using the pen.

\subsubsection{\textbf{DP4.} Avoid explicit interaction modes}

Interaction modes enable an interface to support a wider range of operations. However, constantly switching between modes (e.g., inking vs. gesture) for the same type of input (e.g., pen) can be disruptive to the users' workflow and are known to be a common source of errors~\cite{raskin2000meanings,li2005experimental}.
To avoid explicit interaction modes, we assign semantically meaningful actions to different modalities (e.g., touch for pan/zoom, pen for selection) and leverage a combination of modalities to support advanced variations of simpler actions (e.g., using bimanual pen and touch input to compound selections).

\subsubsection{\textbf{DP5.} Strive for synergy not equivalence}

A common myth about multimodal interaction is that all modes of input can support all operations~\cite{oviatt1999ten}.
Instead of designing specialized interactions (e.g., highly customized and complex gestures or widgets) to ensure equivalence, we support equivalence between modalities only if the equivalence is inherently meaningful.
For instance, we allow binding attributes to encodings using all three modalities (\textit{I1}-\textit{I10} in Table~\ref{tbl:interactions}).
On the other hand, because there is no direct interaction (\textbf{DP2}) to filter marks based on an attribute that is not encoded in the view using pen or touch, we only allow this via speech (e.g., saying ``\textit{Remove movies with an imdb rating under 8}'' when the system shows a scatterplot of \textit{budget} and \textit{gross}).

Furthermore, we also leverage \textit{complementarity-based interactions}~\cite{martin1998tycoon}, where different chunks of information are provided by different modalities and subsequently merged together to perform an operation.
In addition to help accomplish \textbf{DP2} and \textbf{DP3}, complementarity can also facilitate faster interactions and reduce the complexity of speech commands, ultimately improving both the user and system performance~\cite{bolt1980put,vo1993multimodal,martin1998tycoon}.
\edit{For instance, with touch alone, binding multiple attributes to an axis requires multiple interactions with control panel widgets such as dropdown menus (e.g.,~\cite{drucker2013touchviz}).
Alternatively, during \textit{I6:pointing on the axis and speaking}, the axis (target) is implicitly determined by touch whereas speech allows specifying multiple attributes (parameters) as part of the same action.}
\edit{Similarly, to support negative filters, instead of providing additional keep-only button or menu item in touch-only systems (e.g.,~\cite{drucker2013touchviz,sadana2014designing}), a system with \textit{I21:select-and-speak}} can let people select points by drawing a lasso with a pen and saying ``\textit{exclude others}.''
In this case, the target (marks) is specified through the pen while speech provides the operation (via ``\textit{exclude}'') and further modifies the target (via ``\textit{others}'').

\edit{Note that the concepts and interactions in Table~\ref{tbl:interactions} are by no means an exhaustive or definitive set. They are only one sample set of interactions we designed with \textbf{DP1-5} and basic elements of visualization systems in mind.}
In fact, depending on a system's interface, some of these interactions may not even be applicable.
For instance, if a system does not explicitly list attributes as interactive widgets, the \textit{I1:drag-and-drop} and \textit{I2:point-and-tap} interactions involving attribute pills to bind attributes to encodings cannot be used.
However, the \textit{I3:point-and-write} and \textit{I4:point-and-speak} interactions for the same operation remain valid since they rely on the X/Y axes of the chart itself.

\section{InChorus}



\begin{figure}[t]
    \centering
    \includegraphics[width=\linewidth]{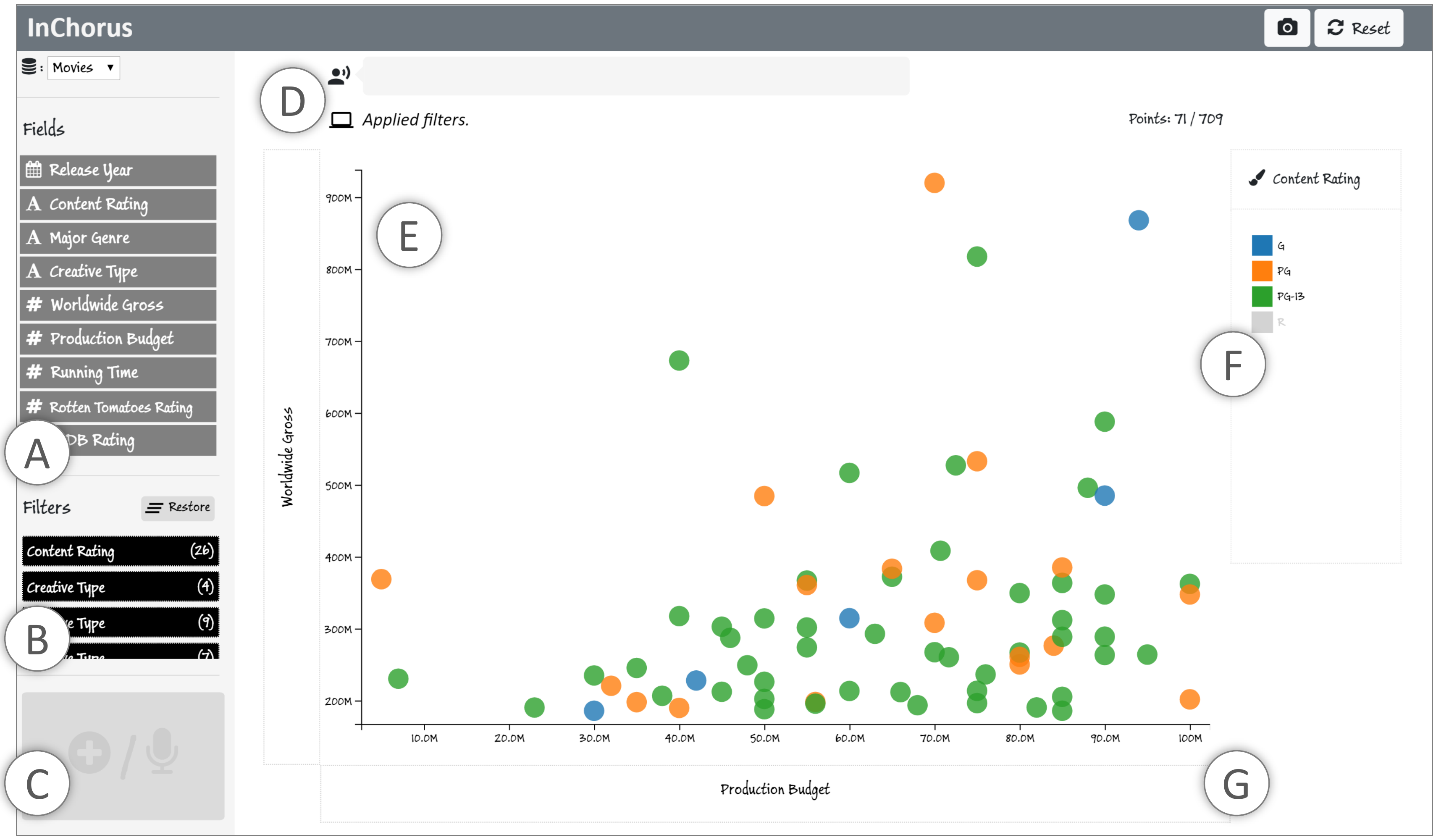}
    \caption{InChorus' interface components. (A) Attribute pills, (B) Active filters, (C) Modifier button, (D) Speech command display and system feedback row, (E) Chart canvas with marks (in this case, circles), (F) Color legend area, and (G) Axis scale and title area.}
    \label{fig:scenario-2-interface}
\end{figure}

To demonstrate how the proposed interactions collectively support visual data exploration on tablets, we employ them in a prototype system, InChorus (Figure~\ref{fig:scenario-2-interface}).
In addition to the basic elements such as axes, legend, attribute pills, etc., similar to previous pen and touch systems (e.g.,~\cite{xia2018dataink,sadana2016expanding}), we also added a modifier button (Figure~\ref{fig:scenario-2-interface}C) that serves two purposes: 1) it allows utilizing bimanual input, which can help avoid explicit mode switches during pen- or touch-only interactions (\textbf{DP4}),
and 2) it serves as a ``record'' button to input voice commands. InChorus uses a ``push-to-talk'' technique: it records speech while a finger is on the modifier button, the X/Y axis title regions, or the color legend title, and executes the recognized command once the finger is lifted.

\begin{figure}[b!]
    \centering
    \includegraphics[width=\linewidth]{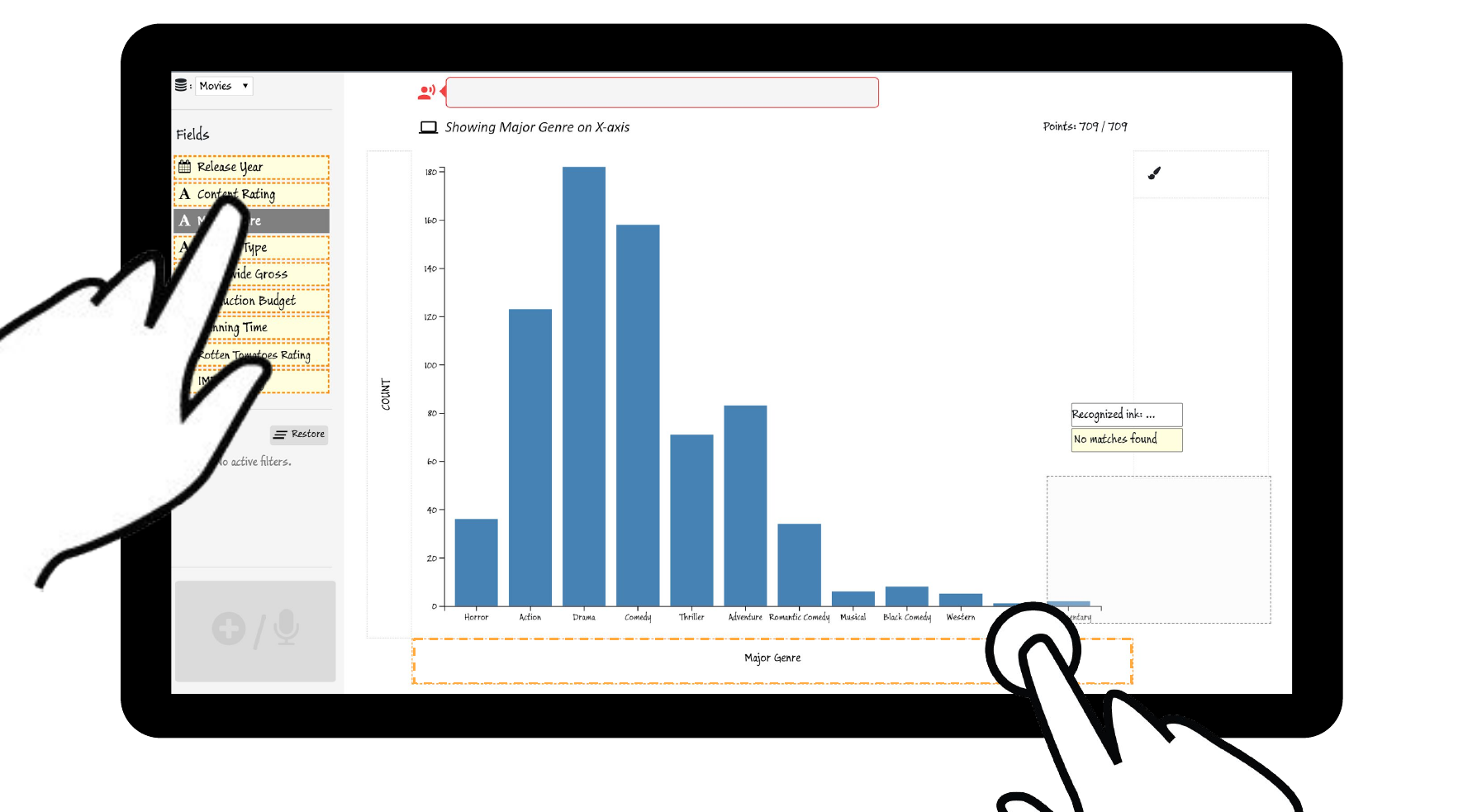}
    \caption{Tapping an attribute while pointing on the x-axis title region binds the data attribute to the x-axis.}
    \label{fig:scenario-0}
\end{figure}

\begin{figure*}[t]
    \centering
    \includegraphics[width=\linewidth]{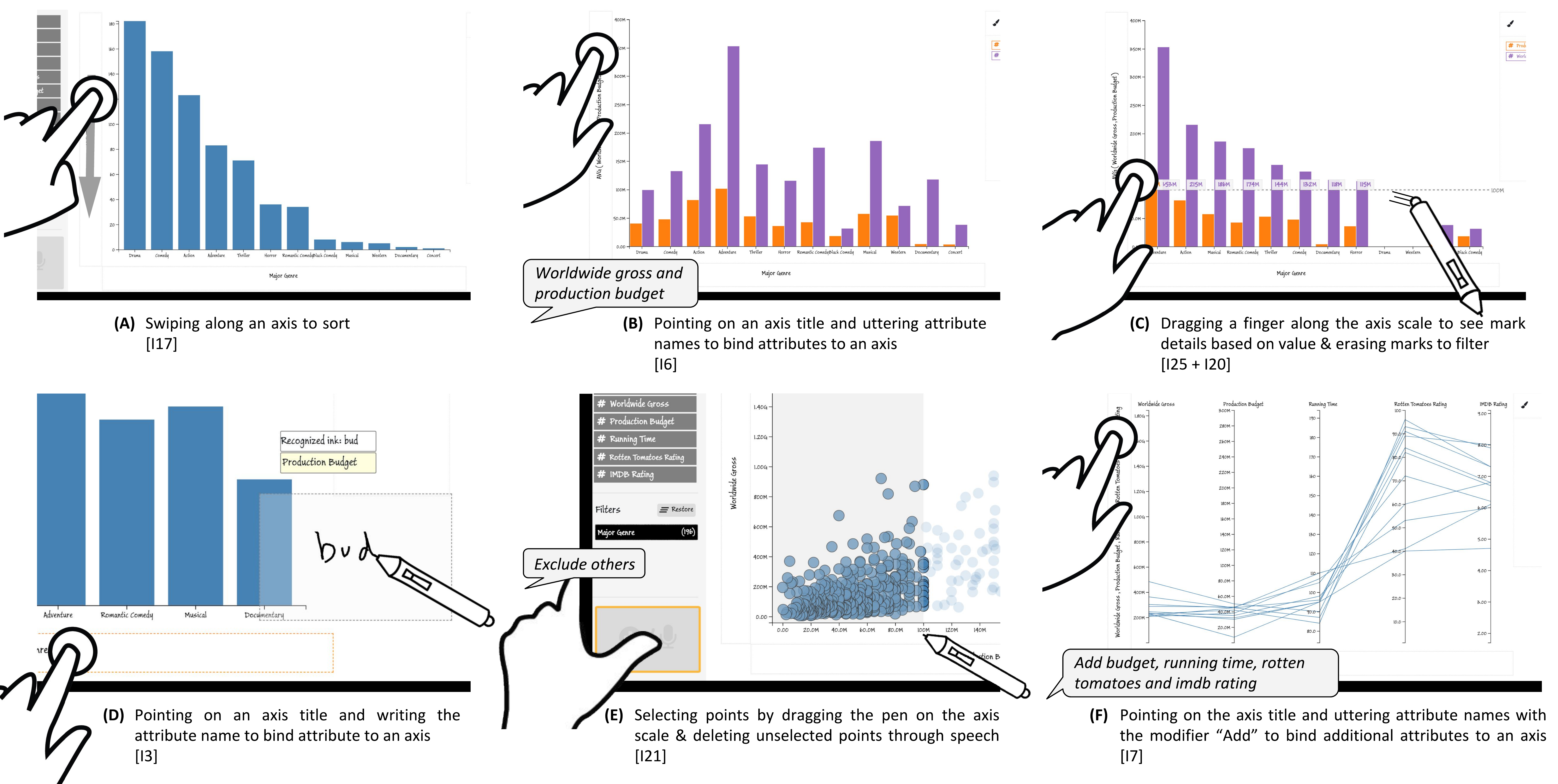}
    \caption{Using InChorus to explore a movies dataset. Sub-figure captions describe the interactions being performed \edit{[along with the corresponding interaction pattern labels from Table~\ref{tbl:interactions}].}}
    \label{fig:scenario-combined}
    \vspace{-1.5em}
\end{figure*}


\subsection{Interacting with InChorus}
\label{sec:scenario}

We now illustrate key interactions in InChorus through a usage scenario.
Imagine that Joe, an analyst at a movie production house, wants to identify movie characteristics his company should focus on for their next investment.
To investigate previously released movies, Joe loads a dataset of 709 movies into InChorus.
The dataset contains nine attributes for each movie, including \emph{Release Year} ({\small{\faCalendar}} temporal), \emph{Worldwide Gross} ({\small{\faHashtag}} quantitative), and \emph{Major Genre} ({\small{\faFont}} categorical), among others; all nine attributes are shown on the left panel (Figure~\ref{fig:scenario-2-interface}A).

\textbf{Identifying key genres.}
To get an overview of values for each attribute, Joe \textit{taps} on individual attributes in the side panel \textit{while pointing} (i.e., holding down a finger) on the X-axis title region (Figure~\ref{fig:scenario-0}).
As he taps through the attributes, InChorus displays univariate summary visualizations (histograms, bar charts, and line charts) based on the attribute's data type.

To see the popularity of different genres, Joe binds the \emph{Major Genre} attribute to the X-axis, creating a bar chart.
Joe decides to sort the bars by the number of movies so he can identify more popular genres faster.
As he \textit{swipes downwards} on the Y-axis, InChorus sorts the genres in descending order by count (Figure~\ref{fig:scenario-combined}A).
To get a sense of the typical return on investment for different genres, Joe adds the gross values to the view by \textit{pointing} on the Y-axis and \textit{saying} ``\textit{Worldwide gross and production budget}.''
This updates the view to a grouped bar chart displaying the average gross and budget for different genres (Figure~\ref{fig:scenario-combined}B).
Now, to look at the highest grossing genres instead of the most popular ones, Joe again wants to sort the view.
However, because two attributes are mapped to the Y-axis, instead of swiping, Joe now \textit{points} on the Y-axis and \textit{says} ``\textit{Sort by worldwide gross in descending order}'' to clearly express his intent.

To see values corresponding to the bars, Joe \textit{drags} his finger along the Y-axis scale.
As he \textit{drags} his finger along the Y-axis scale, InChorus displays a horizontal ruler highlighting the value corresponding to his finger's position and shows details of bars that intersect with the ruler.
Inspecting the chart, Joe decides to only focus on high grossing genres.
He uses the axis value-based ruler as a cut-off point and \textit{erases} bars corresponding to genres with an average \emph{Worldwide Gross} under 100M, filtering them from the view (Figure~\ref{fig:scenario-combined}C).

\textbf{Shortlisting profitable movies.}
With genres shortlisted, Joe now wants to compare the budget and gross for individual movies using a scatterplot.
To do this, he first \textit{erases} the \emph{Production Budget} from the Y-axis title to remove it.
He then \textit{points} on the X-axis title region and starts writing ``\textit{budget}''~in the ink pad, selecting \emph{Production Budget} from the recommended list of attributes (Figure~\ref{fig:scenario-combined}D).
This replaces the \emph{Major Genre} attribute on the X-axis with the \emph{Production Budget}, creating a scatterplot.
Since he works for a relatively small production house, he decides to focus on lower budget movies.
He \textit{drags the pen} along the X-axis scale to select movies with a budget under 100M and \textit{says} ``\textit{exclude others}'' (Figure~\ref{fig:scenario-combined}E) to remove the unselected movies.
To further focus on movies with high a return on investment, he also removes movies with a gross of under 200M.


With the filtered scatterplot, Joe starts examining other attributes.
To understand what types the movies were, Joe maps the \emph{Creative Type} attribute to the color of the points by \textit{saying} ``\textit{Color by creative type}.''
Noticing that \emph{Contemporary Fiction}, \emph{Kids Fiction}, and \emph{Science Fiction} are most popular, he filters out the other movie types by \textit{erasing} them from the legend.
Similarly, mapping the \textit{content rating} to the color of points, Joe removes \textit{R}-rated movies since his company is more interested in movies catering to a universal audience.
This filtering results in the view shown in Figure~\ref{fig:scenario-2-interface}.

Inspecting the scatterplot, Joe notices a set of low budget movies that have made a profit of over 5x.
He selects these movies at the bottom left corner of the view as well as the three highest grossing movies at the top right corner of the view by \textit{holding the modifier button} and \textit{drawing} two free-form lassos around the points using the pen.
Joe then filters out other movies from the chart by \textit{saying} ``\textit{remove others}.''

\textbf{Comparing shortlisted movies.}
Finally, to analyze what characteristics made the shortlisted movies so successful at the box office, Joe wants to inspect and compare the shortlisted movies with respect to the relevant attributes.
Joe first removes the \emph{Production Budget} from the X-axis by \textit{erasing} it.
He then \textit{points} on the Y-axis (currently showing only the \emph{Worldwide Gross}) and \textit{says} ``\textit{Add budget, running time, rotten tomatoes and imdb rating}.''
InChorus, in response, adds the additional attributes to the Y-axis, creating a parallel coordinates plot.
Joe further investigates the shortlisted movies by selecting different value ranges on the parallel axes and identifies a final list of five movies to present as examples of profitable and reliable investments to the management team.

\subsection{Affordances and Feedback}

To make users aware of possible actions, InChorus presents different types of affordances. 
For instance, when the user points on an axis, InChorus highlights attributes that can be mapped to that axis, renders an ink pad to indicate that written input can be provided, and flashes the command display box and microphone red {\small{\textcolor{red}{\faMicrophone}}} to indicate that speech recognition is active (Figure~\ref{fig:scenario-0}).
Note that the attribute pills are contextually highlighted based on the active view and the attribute type.
For example, in Figure~\ref{fig:scenario-0}, \textit{Major Genre} stays dark gray because it is already mapped to the X-axis.
Alternatively, if one pointed on the Y-axis with a categorical attribute shown on the X-axis, the system would only highlight numerical attributes in the panel since we currently do not support visualizations simultaneously showing categorical attributes on both axes.

InChorus also provides constant feedback based on user actions.
In addition to visual feedback for direct pen/touch actions (e.g., fading unselected points in Figure~\ref{fig:scenario-combined}E, adding \edit{an orange} stroke when pointing on an axis title as shown in Figure~\ref{fig:scenario-0}), the system also displays three types of textual feedback messages above the chart area (Figure~\ref{fig:scenario-2-interface}D): 1) Success: when the system successfully executes operations in response to user actions (example messages include~\textit{Coloring by Major Genre}; \textit{Sorted bars by Worldwide Gross in descending order}), 2) Void action: when users performs a valid operation but the operation has no effect on the view (example messages include~\textit{No points meet that filtering criteria}; \textit{Bars are already sorted in descending order by IMDB Rating}), and 3) Error: when users perform invalid actions or the system is unable to interpret a speech command (example messages include~\textit{The pen cannot be used in the panel area. Please use touch.}; \textit{Unable to process that command. Please try a different one}).

By providing contextually-relevant affordances before an action and complementing them with feedback after actions, InChorus helps users both know what actions are available as well as interpret the system's reactions to their actions.

\subsection{Implementation}


InChorus is implemented in JavaScript as a web-based application.
All visualizations are rendered using D3.js~\cite{bostock2011d3}.
Pen and touch inputs are collected as standard JavaScript events and processed by custom event handlers.
InChorus uses the HTML5 speech recognition API~\cite{webspeechapi} for translating speech-to-text.
To improve recognition accuracy, the speech recognizer is trained with the operation-specific keywords (Table~\ref{tbl:interactions}) and attributes and values in the loaded dataset.

We implemented a custom JavaScript-based lexical parser to interpret the recorded NL commands.
The lexicon consists of the attributes and values in the dataset as well as manually defined operation-specific keywords.
To identify operations, targets, and parameters, the system compares the tokenized input string to the lexicon.
If it is unable to detect a target using the NL command alone, the system employs multimodal fusion and infers the target through the invoking instrument (e.g., axes) and the active view (e.g., selected points).
\section{User Study}

We used InChorus as a test bed to assess the general usability of the proposed interactions and gather subjective feedback.
In particular, since multimodal interaction with visualizations through pen, touch, and speech is a novel concept, we wanted to assess its practical viability and see whether people actually adapt to using this style of interaction and are able to perform common visual analysis tasks.

\subsection{Participants and Setup}
We recruited 12 participants (P1-P12; six females, five males, and one ``undisclosed''), ages 27-55, via email through mailing lists of a large technology company.
All participants rated themselves as being fluent English speakers and had a working knowledge of data visualizations (i.e., understood basic visualization types and elements such as axes, legends) but not necessarily specific visualization systems (e.g., Tableau, Microsoft Power BI).
In terms of prior experience with input modalities, all participants said they use touch-based systems on a daily basis.
Two participants said they use a pen once in a few weeks, five said they had used it in the past but do not use it regularly, and five said they had no experience working with a pen.
Eight participants said they use voice-based systems on a daily basis, three said they use it on a weekly basis, and one said she only occasionally uses voice-based systems.

Sessions were conducted in-person in a quiet conference room.
Participants interacted with InChorus on Google's Chrome browser on a 12.3'' Microsoft Surface Pro set to a resolution of 2736~x~1824.
Participants were encouraged to position the device in a way that was most comfortable for them.
A 24'' monitor was used to display the study instructions and tasks as a slide show.
Participants received a \$50 gift card as a compensation for their time.
All sessions were audio and video recorded.

\subsection{Procedure and Tasks}
Each study session had four phases including a training phase, two task phases, and debriefing.
The study protocol and tasks were iteratively refined through 12 pilot sessions.
We included two types of tasks to emulate two significantly different visual analysis scenarios.
Specifically, the first task phase emulates scenarios where users know the operations they want to perform and have to communicate that intent to the system through interactions.
On the other hand, the second task phase emulates scenarios where users first need to think about the task they want to accomplish (e.g., think about the attributes they want to use and the type of chart they want to create), translate that task into system operations (e.g., binding attributes to specific encodings), and finally perform those operations through the supported interactions.
Each session lasted between 71-124 minutes (average: 86 min).

\textbf{Introduction \& Training.}
After they provided consent and filled out a background questionnaire, participants were introduced to the various system operations along with possible interactions for each operation using a dataset of 303 cars with eight attributes (e.g., \textit{Horsepower}, \textit{Acceleration}, \textit{Origin}) for each car.
During this phase, as training, participants were free to practice the interactions until they felt confident performing them.
This phase lasted approximately between 37-60 minutes (average: 46 min).

\textbf{Task Phase 1: Replication and Value Identification.}
In this phase, participants were given four tasks using the IMDB movies dataset introduced as part of the usage scenario earlier. Each task consisted of a set of one to four sub-tasks that displayed a visual state for participants to replicate or values they had to identify. 
For example, one of the tasks had four sub-tasks requiring participants to 1) recreate a given grouped bar chart, 2) filter out two categories of values, 3) switch to a multi-series line chart, and 4) identify series values for a specific year. 
This first task phase took approximately 8-26 minutes (average: 13 min) to complete.

\textbf{Task Phase 2: Fact Verification.}
Participants were given a dataset of 500 US colleges with nine attributes for each college including a number of numerical (e.g., \textit{Cost}, \textit{Admission Rate}) and categorical attributes (e.g., \textit{Control Type}, \textit{Region}).
Following the jeopardy-style evaluation~\cite{gao2015datatone} for visualization NLIs, we gave participants five statements (e.g., \textit{There are more public schools in Southwest than the Great Lakes}) that they had to mark as true or false based on their exploration of the data.
As with other systems evaluated using this methodology~\cite{gao2015datatone,srinivasan2018orko}, parroting the given statement to the system would not result in the answer.
In addition to stating the answer, participants also had to verbally justify their responses and take screenshots of visualizations they used.
This phase lasted approximately between 7-28 minutes (average: 14 min).

\textbf{Debrief.}
At the end of the session, we had a debriefing phase that included a post-session questionnaire consisting of likert-scale questions regarding their interaction experience and an interview.
During the interview, we asked participants general questions about their overall experience, asking them to list interactions they particularly liked/disliked, as well as targeted questions based on our observations during the session.

\begin{figure*}[t]
    \centering
    \includegraphics[width=.95\linewidth]{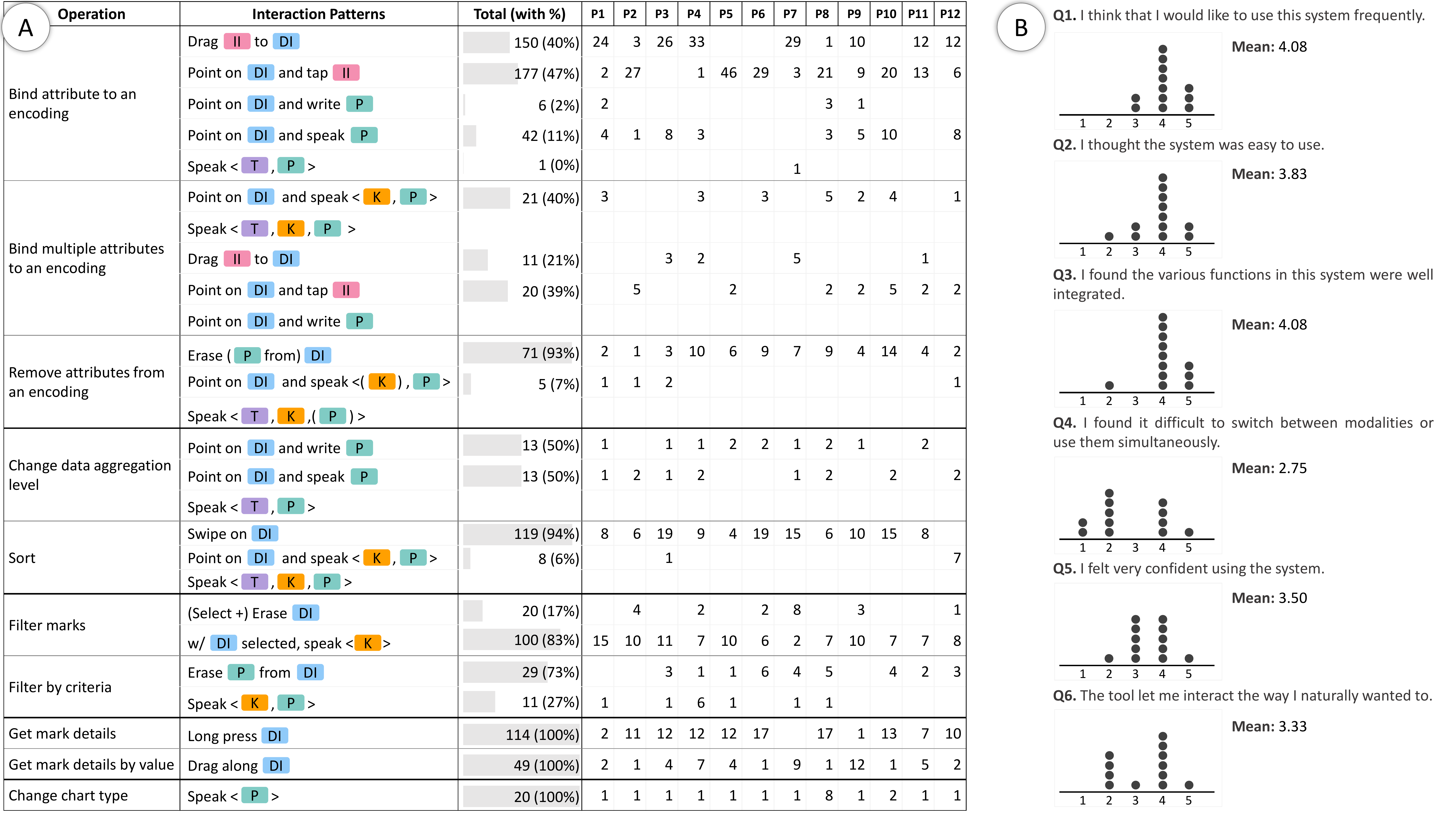}
    \caption{
    Summary of study results.  
    (A) Operations executed during the study along with the frequency of interaction patterns which were used to perform those operations.
    Bar widths are normalized for individual operations to facilitate comparison of alternative interaction patterns
    and 
    (B) Responses to post-session likert-scale questions about the interaction experience.
    \vspace{-1em}
    }
    \label{fig:study-results}
\end{figure*}


\subsection{Results}

All participants successfully replicated charts or identified values for the four tasks in the first phase.
For the second task phase, nine participants correctly verified all five statements whereas three (P1, P5, P10) correctly verified three out of the five statements.
The average completion time for individual tasks was 3 minutes for the first phase and 2:30 minutes for the second.

\subsubsection{Interaction Summary}

To track participants' interactions, we reviewed the session videos to count the number of times participants attempted operations in Table~\ref{tbl:interactions} along with which interaction patterns they used.
We identified a total of 1197 attempts, out of which 1000 (83\%) executed successfully, 67 (6\%) were invalid operations (e.g., mapping an attribute to color with no attribute mapped to the X/Y axes), 15 (1\%) used unsupported speech commands (e.g., invoking an axis-value based ruler and saying ``\textit{Remove points under this}'') or pen/touch gestures (e.g., dragging an attribute from the color legend to the X axis), and 115 (10\%) involved erroneous interactions \edit{(e.g., speech recognition errors, conflicting pen and touch).}
\remove{These errors included dragging an attribute pill outside the screen or dropping it outside the axes/legend (36), speech recognition errors (32), forgetting to trigger recording of voice commands before speaking (16), conflicting pen and touch (e.g., trying to drag an attribute pill with the pen) (15), dragging an attribute while pointing on an axis (6), forgetting to select from the list of recognized items while writing (5), and issuing incomplete speech-only commands (e.g., saying ``\textit{Remove under 1200}'' without specifying an attribute name) (5).}


For invalid operations, unsupported and erroneous interactions, participants often reattempted their initial interaction one or more times before switching to a different interaction pattern or operation.
Hence, to avoid double counting interaction patterns by including these reattempts, we only summarize the 1000 valid and successful interactions in Figure~\ref{fig:study-results}A.

\subsubsection{Subjective Feedback}

Figure~\ref{fig:study-results}B summarizes participants' responses to the post-session questionnaire.
In general, participants were positive about the overall experience stating they found the system functionalities well integrated (\textit{M} = 4.08 out of 5) and that they would want to use the system frequently (\textit{M} = 4.08).
However, all participants noted that there was a learning curve especially during the training phase.
That said, participants felt the interactions were intuitive and just needed some getting used to as they had no prior experience with multimodal interfaces.
For instance,\remove{P12 said ``}\textit{\remove{It's a lot to take in in a short amount of time but it felt reasonably intuitive.''}}
P6\remove{ even} compared her initial reaction to using a new type of keyboard and said ``\textit{...it was a cohesive system. It's just kind of getting the muscle memory down. It's like when you switch to someone else's keyboard you know where everything is but you just can't type.}''
However, after the training, not only did she successfully complete all tasks but as highlighted by the counts in Figure~\ref{fig:study-results}A, did so using a variety of patterns including unimodal interactions, bimanual interactions, and multimodal interactions combining pen/touch with speech.

While seven participants said they did not find it difficult to switch between or combine modalities, five said it was confusing, in particular to differentiate between pen and touch.
For instance, P9 said ``\textit{Distinguishing and switching between voice versus not-voice was not difficult but pen versus finger was a tougher one. I don't think I make as much of a distinction between pen and finger and only use the pen to be more specific, it's a finer point as opposed to being a different tool}.''
We believe this confusion may have been exacerbated by the fact that the three participants giving this feedback had never used a pen before and the remaining two had used it minimally in the past.
Nonetheless, this confusion fuels the open challenge with designing bimanual pen and touch interaction~\cite{frisch2009investigating,hinckley2010pen+}, deeming further investigation.

Lastly, on average, participants were neutral about how ``natural'' it felt to interact with the system (\textit{M} = 3.33), with four participants stating they felt the system did not support their natural workflow.
During the interviews, we noted that more so than the interaction patterns, this stems from the fact that some participants preferred not to manually specify mappings to explore the data.
For instance, P5 said ``\textit{I'm an excel person. If I had different templates I would have selected different types of graphs and seen what the data looks like and then chose the scatter graph.}'' 

\section{Discussion and Future Work}


\subsection{Towards a Grammar for Multimodal Interaction}

Although we focus on tablets as our primary use case in this paper, the underlying conceptual framework (designing and describing multimodal interactions with visualizations in terms of operations, targets, parameters, and instruments) is not limited to tablets or the listed operations.
For instance, the \textit{I3:point-and-write} or \textit{I4:point-and-speak} interactions with X/Y axes or the color legend to bind attributes to encodings can be employed as-is in a system like SketchInsight~\cite{lee2015sketchinsight} that supports visual data exploration on interactive whiteboards.

In addition to different form-factors, the presented concepts can also be used to design consistent interactions for more advanced visual analysis operations.
One such operation might be generalized selection~\cite{heer2008generalized} that allows people to specify selection criteria relative to a subset of the data they are interested in.
Previous touch-only systems have supported generalized selection through explicit interaction modes and WIMP-style widgets~\cite{sadana2016expanding}.
However, multimodal systems could support this in a more fluid manner by allowing users to specify their selections via pen/touch and generalization criteria via speech.
For instance, we can define a new interaction pattern: \textit{Select \symbolDI~and speak <\symbolK, \symbolP>} where \symbolDI: marks on the view, \symbolP: selection criteria in terms of attributes and values, and \symbolK: [``select'', ``highlight'', ``lower'', ...].
Thus, based on the user selecting a \egDI{point} in a scatterplot and saying ``\textit{\egK{Select} \egP{Action} movies with a \egK{lower} \egP{budget} than this},'' the system can infer the operation and leverage the active selection to find and select desired points, all while preserving the current interactions in Table~\ref{tbl:interactions}.

While these are just examples, in general, standardizing the design of interactions with respect to logical concepts such as operations, targets, parameters, and instruments presents a compelling opportunity to create a unifying grammar to describe multimodal interactions with visualization systems.
In addition to streamlining multimodal interaction design, such a grammar could also help develop extensible toolkits and interpreters that let designers create, share, and modify interaction patterns across visualization types and interfaces.

\subsection{Potential of ``Restricted'' Natural Language Interfaces}

A majority of the work on visualization NLIs has focused on developing interpretation techniques for complex and underspecified commands~\cite{gao2015datatone,setlur2016eviza,hoque2018applying,setlur2019inferencing}.
While this is the holy grail of NLIs in general, responding to high-level questions and underspecified commands is challenging and highly error prone due to issues such as ambiguity and preserving context across commands.
Our work sheds light on a more modest but less error prone class of NLIs for visualization.
Specifically, with InChorus, we allow users to issue short keyword-based commands that can be used in conjunction with another modality (\textbf{DP5}) or individually to perform low-level operations.
We found that these simple commands allowed participants to perform required operations while improving the overall speech recognition and interpretation accuracy:
out of a total of 274 utterances, we only had 32 (11\%) recognition errors and 7 interpretation errors (2\%).
This speech recognition error rate of an average of 3\% per session is noticeably lower than previous speech based visualization systems like Orko~\cite{srinivasan2018orko} (avg.~of 16\% per session).
With such results in mind, in line with recent work advocating for ``restricted'' NLIs in complex domains~\cite{mu2019we}, perhaps an opportunity lies in exploring ``restricted'' NLIs for visualization that build upon simple lexicons and smoothen the transition from current direct manipulation and WIMP-based systems to NLIs.
Although they start simple, over time, such systems could evolve to support more complex commands, incrementally exposing users to more advanced system functionality~\cite{furqan2017learnability,srinivasan2019discovering}.

\subsection{Synergy between Input Modalities}
Our goal was to enable an overall consistent and fluid interaction experience during visual analysis, accommodating users' individual preferences. To this end, instead of aiming to achieve equivalence between modalities, we synergistically combined three input modalities (i.e., pen, touch, and speech), leveraging their unique advantages (\textbf{DP5}).
The distribution of interaction frequencies in Figure~\ref{fig:study-results}A illustrates that the proposed multimodal interactions accommodated varied user preferences while allowing all participants to perform common visual analysis tasks.
While some participants (e.g., P1, P8, P9) were open to trying a wider range of interactions and using all modalities and others (e.g., P5, P6, P11) preferred resorting to touch as much as possible, all participants adapted themselves to using multimodal input and, when needed, successfully used different modalities (individually and in combination) to complete tasks.
This was particularly encouraging given that some participants had no experience with some modalities (e.g., five out of 12 participants had never used a pen).
P7 aptly summarized his overall experience of interacting with InChorus stating ``\textit{It [InChorus] feels completely integrated like this is one thing it's not like this is the pen stuff that I'm doing and now I got to sort through the finger things I can do or the things with voice commands it was like I'm going to interact with this system however best suits me and I'm able to do that nine times out of ten}.''


\subsection{\edit{Improving Affordances and Feedback}}

\edit{From the 115 erroneous interactions, we identified seven types of errors that participants encountered---\textit{E1:} dragging an attribute pill outside the screen or dropping it outside the axes/legend (36), \textit{E2:} speech recognition errors (32), \textit{E3:} forgetting to trigger recording of voice commands before speaking (16), \textit{E4:} conflicting pen and touch (e.g., trying to drag an attribute pill with the pen) (15), \textit{E5:} dragging an attribute while pointing on an axis (6), \textit{E6:} forgetting to select from the list of recognized items while writing (5), and \textit{E7:} issuing incomplete speech-only commands (e.g., saying ``\textit{Remove under 1200}'' without specifying an attribute name) (5).
While a majority of these errors were primarily due to inexperience with the interface \& input modality (E1, E5, E6) or technological errors (E2), others (E3, E4, E7) were more specific to multimodal input.
In addition to the errors, there were also 67/1197 (6\%) instances where participants performed a valid interaction but the intended operation was invalid given the state of the view (e.g., sorting a numerical axis in a scatterplot).}

\edit{As described earlier, we had designed affordances and feedback mechanisms to prevent such errors and invalid operations and help users recover from them.
However, during the study, participants rarely noticed the affordances and feedback, often repeating the same interactions multiple times before realizing it was an error or an invalid operation.
Going forward, it would be interesting to investigate alternative mechanisms such as multimodal feedback (e.g.,~auditory and haptic feedback), automated voice recording (e.g.,~automatically triggering recording after selections are made), and visual aids to clarify support for pen versus touch input (e.g.,~different background colors or animation overlays for the first time interactions).
}
\remove{Apart from the errors, there were also 67/1197 (6\%) instances where participants performed a valid interaction but the intended operation was invalid given the state of the view.
Examples of invalid operations included binding an attribute to color when there is no attribute mapped to the X/Y axis or binding categorical attributes to both X and Y axes (not supported by the current visualization types).
As mentioned earlier, we included affordances and feedback mechanisms to prevent or help users recover from such scenarios.
However, during the study, participants rarely noticed the feedback, often repeating interactions multiple times before realizing it was an invalid operation.
This highlights another challenge that warrants further research on \textit{how to provide noticeable yet unobtrusive feedback} during multimodal interaction.
Perhaps a unique opportunity for this lies in exploring multimodal fission-based feedback techniques that provide both visual and non-visual (e.g., auditory or haptic) feedback.}

\subsection{\edit{Balancing Novelty and Familiarity}}

The high frequency of drag-and-drop interactions in both Figure~\ref{fig:study-results}A and the list of erroneous interactions raises an interesting question about why did participants not switch to alternative patterns upon encountering errors.
On further inspection, we found that while most participants switched to the \textit{I2:point-and-tap} interaction, there were some (P3, P4, P7) who persisted with \textit{I1:drag-and-drop} largely due to its familiarity.
For instance, P4 had 14 drag-and-drop errors but still preferred it over other patterns for binding attributes to encodings.
When asked about why he did not switch to a different pattern, he initially said ``\textit{drag-and-drop feels quicker even if I fail than stop and try and articulate the word to match the AI to get what I want out of it.}''
He later acknowledged that point-and-tap would have been more accurate and possibly faster, ultimately attributing his preference for drag-and-drop to familiarity.\remove{saying} \remove{``}\textit{\remove{it's just the historical effect of my brain working with having done so many things in the past with drag-and-drop.}\remove{''}}
Such observations further motivate the need to explore interfaces that give users the freedom to choose their preferred style of interaction and balance between familiarity and novelty depending on the task at hand (e.g., if speed and efficiency were paramount for the task, P14 may have preferred \textit{I2:point-and-tap} over \textit{I1:drag-and-drop}).

\subsection{Study and Prototype Limitations}


The user study helped us verify that participants were able to successfully adapt and use the proposed multimodal interactions consistently across different visualization types.
However, as with most lab studies, our study has some practical limitations and was scoped to a predefined set of tasks, datasets\edit{, and visualizations}.
Changing either of these could affect how users behaved with the system and the inferences made.
For instance, in his interview, P9 said ``\textit{When there's only a handful of fields here it's easy to drag and select but if there are like hundreds of fields, that's when you would see me writing the name of the field I'm looking for.}'' suggesting that his interaction behavior would be different for datasets with a larger number of attributes.
Thus, an important next step is to leverage the feedback from this initial study and test the interactions in the context of additional types of datasets \edit{and visualizations (e.g., 
heatmaps, node-link diagrams)}, perhaps also investigating more open-ended exploratory tasks.
\edit{Furthermore, because our study focused on adoption and usability of interactions, follow-up experiments are required to more deeply understand the use of multiple modalities along with the associated cognitive challenges and analytical benefits.}

Although InChorus was primarily designed to illustrate and test the proposed interactions, the system can be improved and expanded to enhance the overall usability and user experience of interacting multimodally.
Some of these improvements include adding an undo/redo feature, providing more details about filtered points, and adding visualization types that support additional configurations of the view (e.g., heatmaps to allow categorical attributes on both axes).


\section{Conclusion}

We present multimodal interactions for performing visual data analysis operations on tablet devices, designed with the high-level goal of maintaining interaction consistency across different types of visualizations.
We describe the design process we followed to systematically develop multimodal interactions by modeling interactions with respect to core concepts including \textit{operations}, \textit{targets}, \textit{parameters}, and \textit{instruments}.
Through a user study with 12 participants performing visual analysis tasks with a prototype system implementing the proposed interactions, we discuss how participants adapted to using multimodal interaction and how the freedom of expression afforded by multiple modalities accommodated their interaction preferences.
Ultimately, by highlighting the potential benefits of multimodal input and promoting the systematic development of multimodal interactions, we hope to encourage the design of a new generation of visualization systems that can facilitate more natural and fluid human-data interaction, accommodating varying user preferences.

\balance{}

\bibliographystyle{SIGCHI-Reference-Format}
\bibliography{inchorus}

\end{document}